\documentclass{article}

\usepackage[nonatbib, final]{neurips_mlsb_2020}

\usepackage[utf8]{inputenc} % allow utf-8 input
\usepackage[T1]{fontenc}    % use 8-bit T1 fonts
\usepackage{hyperref}       % hyperlinks
\usepackage{url}            % simple URL typesetting
\usepackage{booktabs}       % professional-quality tables
\usepackage{amsfonts}       % blackboard math symbols
\usepackage{nicefrac}       % compact symbols for 1/2, etc.
\usepackage{microtype}      % microtypography
\usepackage{titlesec}
\usepackage{neurips_mlsb_2020}
\usepackage{verbatim}
\usepackage{graphicx}
\usepackage{amssymb}

\usepackage{caption} 
\usepackage{amsmath}
\usepackage{bm}
\usepackage{upgreek}

\DeclareMathOperator*{\argmin}{argmin}

\usepackage{siunitx}
\usepackage[ruled,vlined]{algorithm2e}

\usepackage{tabularx}
\usepackage{floatrow}
\newfloatcommand{capbtabbox}{table}[][\FBwidth]

\usepackage{wrapfig}

\title{Learning a Continuous Representation of 3D Molecular Structures with Deep Generative Models}

\author{%
  Matthew~Ragoza\thanks{These authors contributed equally to this work} \\
  Comp. \& Systems Biology \\
  University of Pittsburgh \\
  Pittsburgh, PA 15213 \\
  \texttt{mtr22@pitt.edu} \\
  \And
  Tomohide~Masuda\footnotemark[1] \\
  Comp. \& Systems Biology \\
  University of Pittsburgh \\
  Pittsburgh, PA 15213 \\
  \texttt{tmasuda@pitt.edu} \\
  \And
  David Ryan Koes \\
  Comp. \& Systems Biology \\
  University of Pittsburgh \\
  Pittsburgh, PA 15213 \\
  \texttt{dkoes@pitt.edu} \\
}

\titlespacing{\section}{0pt}{*1.5}{*.5}
\titlespacing{\subsection}{0pt}{*1.5}{*.5}
\titlespacing{\subsubsection}{0pt}{*1.5}{*.5}
\titlespacing{\paragraph}{0pt}{*2}{*1}

\begin{document}

\maketitle

\begin{abstract}
  Machine learning in drug discovery has been focused on virtual screening of molecular libraries using discriminative models. Generative models are an entirely different approach that learn to represent and optimize molecules in a continuous latent space. These methods have been increasingly successful at generating two dimensional molecules as SMILES strings and molecular graphs. In this work, we describe deep generative models of three dimensional molecular structures using atomic density grids and a novel fitting algorithm for converting continuous grids to discrete molecular structures. Our models jointly represent drug-like molecules and their conformations in a latent space that can be explored through interpolation. We are also able to sample diverse sets of molecules based on a given input compound and increase the probability of creating valid, drug-like molecules.
\end{abstract}

\section{Introduction}

The goal of drug discovery is to find novel molecules that bind to target proteins of pharmacological interest. This involves a long, costly pipeline in which a massive chemical space is repeatedly filtered into the subset of compounds most likely to be active. Computational methods promise to streamline this process through rapid scoring and ranking of large libraries of molecules \textit{in silico} prior to confirming their activity with experimental assays \cite{Seifert2003}. This approach, called virtual screening, has established a significant presence in the modern drug development toolkit.

Despite its utility, virtual screening has inherent limitations. Though faster and cheaper than high-throughput experimental screening, it cannot propose compounds absent from existing libraries or suggest changes to input compounds that improve their desired properties. The need to explicitly enumerate molecules for screening means that even sub-linear time virtual screening algorithms \cite{sunseri2016pharmit} cannot cover the entirety of chemical space.

An alternative class of machine learning methods called generative models could overcome these limitations. Virtual screening falls within the category of discriminative models, which are designed to predict some desired property, such as binding affinity, conditioned on the input data. In a generative modeling formulation, the goal instead becomes learning  to directly sample from the underlying distribution of drug-like molecules in order to yield novel active compounds.

In computational drug design, generative models have thus far been used to map from two-dimensional molecules to a continuous latent space that is then explored to produce molecules with desired properties. Here we describe for the first time a deep learning model that represents and generates constitutionally diverse three-dimensional molecular structures. We encode 3D molecules to a latent space using atomic density grids as input and decode them in the same format. Then we apply a novel optimization algorithm to fit 3D chemical structures to generated atomic densities and reliably produce valid molecules.

\section{Related Work}

% Overview of machine learning for drug discovery

In the last decade, major advances in computer vision \cite{krizhevsky2012imagenet, lecun2015deep} have sparked a surge of interest in deep learning for drug discovery \cite{angermueller2016dl}. Researchers have applied deep learning to binding discrimination for virtual screening \cite{wallach2015atomnet, ragoza2017cnn} and pose ranking for molecular docking \cite{ragoza2017cnn} by using atomic density grids to treat classification and regression of molecular structures as analogous to three-dimensional image recognition. Other work has trained deep neural networks for binding affinity prediction \cite{gomes2017, jimenez2018kdeep} and quantum energy estimation \cite{schutt2017} using atomic coordinate-based input representations.

% Deep generative models

Deep learning for generative modeling generally falls into two frameworks. In a variational autoencoder (VAE) \cite{kingma2014vae}, an encoder and decoder network are trained to map between the input space and a set of latent random variables that follow a predetermined distribution. This allows the model to sample from the prior and create novel data samples. Generative adversarial networks (GANs) \cite{goodfellow2014gan} are a complementary approach \cite{larsen2016vaegan} that concurrently train a discriminative network to distinguish real and fake data, and a generative network to create samples that the discriminator classifies as real. GANs have proven successful at learning to generate novel image-like data. \cite{radford2016dcgan}

% Generating SMILES strings

The first efforts applying deep generative models to molecule generation \cite{gomez-bombarelli2016, segler2017lm} used the SMILES string format \cite{weininger1988smiles}. Treating chemicals as linguistic sequences allows the use of recurrent networks that have been effective at modeling natural language. However, similar molecules can have very different SMILES strings due to their lack of permutation invariance, and invalid SMILES strings can be produced unless grammatical constraints are imposed on the model \cite{kusner2017gramvae,dai2018sdvae} or valid strings are rewarded through reinforcement learning \cite{guimaraes2018organ}. Lastly, SMILES strings only describe molecules in terms of their connectivity, so they cannot represent their 3D spatial conformations.

% Generating molecular graphs

Others have leveraged graph convolutions \cite{kearnes2016graphconv} to train deep generative models on molecular graphs \cite{simonovsky2018graphvae,decao2018molgan,jin2019jtvae,kwon2019graph}. These more naturally capture many invariances present in chemical structures, but require expensive or approximate graph matching algorithms to compare output to input molecules. Molecular graphs also suffer from the possibility of invalid outputs, so reinforcement learning has been used with them as well \cite{decao2018molgan,kwon2019graph}. Most work with molecular graphs has been confined to two dimensions, though forays have been made into generating them along with distance matrices \cite{gebauer2018distmat,hoffmann2019distmat}. These 3D molecular graph-based models are promising, but have so far only been evaluated at generating different constitutional isomers of a single, fixed set of atom types.

% Generating atomic density grids

 Atomic density grids have several advantages over other molecular representations, as well as limits. Grids can process as many atoms as can fit within their bounds in constant time and memory, while SMILES strings and molecular graphs scale linearly and quadratically with the number of atoms. Grids facilitate learning from holistic 3D spatial configurations with convolutions \cite{ragoza2017cnn} and are permutation invariant, though not rotation invariant. Finally, density grids must be converted into a discrete representation to generate valid molecules. Previous efforts at generating atomic densities used a recurrent captioning network to convert them to SMILES strings \cite{skalic2019shapevae}. However, this relinquishes the 3D nature of the generative model. Rather than construct 2D molecules from 3D densities, we solve for discrete 3D molecules that best fit atomic densities through optimization.

\iffalse
other possibly relevant work:

 empirical/ML virtual screening functions
\cite{Williams2008, Bohm1994score, Wang1998, Wang2002xscore, trott2010, sminapaper}
\cite{Chen2007, Ballester2010prosp, zilian2013sfcscore, durrant2010nnscore, durrant2011nnscore, Ashtawy2015}

atom fitting similar to multiple object detection/localization in images

neural fingerprints \cite{duvenaud2015}

graph convolutions \cite{kearnes2016graphconv}

MOSES benchmark

NeVAE dgn for molecular graphs

Mol-CycleGAN

3D fingerprints \cite{axen20173dfp}

neural fingerprints

3d shape based ligand VS \cite{Sastry2011}

shape VS using gaussians \cite{Proschak2008}

2d fp vs 3d shape vs \cite{hu2012performance}

guass docking funcs \cite{McGann2002}

pubchem \cite{Bolton2008}

pubchem \cite{wang2009}

caffe \cite{jia2014caffe}
\fi

\section{Methods}

\subsection{Representing molecules as atomic density grids}

We represent molecules in a grid format amenable to convolutional network training that was first described in \cite{ragoza2017cnn}. To convert molecules to grids, we assign each atom a type based on its element, aromaticity, hydrophobicity, and hydrogen donor/acceptor state. Then, the atoms are represented as continuous, Gaussian-like densities on a 3D grid with separate channels for each atom type, akin to an image with separate channels for each color. We compute the values in each atom type channel by summing over the density of each atom with the corresponding type at each grid point. Grids span a cube of side length 23.5 \si{\angstrom} with 0.5 \si{\angstrom} resolution, resulting in $48 \times 48 \times 48$ points per channel. We center the grids on the input molecule and compensate for the lack of rotation invariance by randomly rotating during training. We also evaluate 10 rotations of each molecule in the validation and test sets. This is facilitated by computing grids on-the-fly using \texttt{libmolgrid}, an open-source, GPU-accelerated molecular gridding library \cite{sunseri2020libmolgrid}.

% TODO Atom typing scheme in supplemental materials

\subsection{Constructing molecules from atomic density grids}

The inverse problem of converting density grids back into discrete 3D molecular structures does not have an analytic solution, and instead must be solved through optimization. \texttt{libmolgrid} allows us to compute the grid representation of a partial atomic structure and backpropagate a gradient from grid values to atomic coordinates. We can also estimate the initial locations of atoms on a density grid by selecting from the grid points with the largest density values. Therefore, we devised a solver algorithm that combines atom detection, gradient descent, and beam search (in which the current top-$k$ structures are stored and expanded) to find a set of atoms that best fits a reference density.

This algorithm returns atom types and coordinates, but further processing is required to produce a valid molecule. We add bond information based on the returned atom types and geometry with customized functions from OpenBabel \cite{oboyle2011openbabel, openbabel} that infer connectivity and bond order while respecting the constraints of our atom typing scheme. These constraints include aromatic ring membership, presence of bonds to hydrogen and heteroatoms, and formal charge. The validity of the resulting molecules are assessed using RDKit--a molecule is valid if and only if it is composed of a single, connected fragment and raises no errors when passed to the RDKit function \texttt{SanitizeMol} \cite{rdkit}.

\subsection{Generative model architectures and training}

We trained deep generative models using atomic density grids as both input and output. Generators were trained as convolutional autoencoders to reconstruct their input by minimizing the squared error, or L2 loss, of the output density with respect to the input density. In addition, we trained our models as GANs, using an adversarial discriminative network to encourage the generated densities to match the overall distribution of real densities. Initial experiments showed that including this GAN loss simultaneously helped to reduce L2 loss and encourage the model to generate density for less common atom types.

Both standard (AE) and variational (VAE) autoencoders were trained as the generator component. Their architectures were identical except that the VAE had a probabilistic latent space that was trained with a Kullback-Liebler (KL) divergence loss function to follow a standard normal distribution. The VAE combined its KL divergence, L2, and adversarial loss functions with weights of 0.1, 1.0, and 10.0 respectively, while the AE weighted its L2 and GAN losses equally at 1.0. Models were trained using the Adam optimization algorithm \cite{kingma2019adam} with hyperparameters $\alpha = 1e-5, \beta_1 = 0.9, \beta_2 = 0.999$ and a fixed learning rate policy for 100,000 iterations, using a batch size of 16. The full details of our model architectures are found in Figure 7.

\subsection{Data sets}

In order to train deep neural networks to generate molecules in 3D, we downloaded the full set of commercially available stock compounds from the MolPort database \cite{molport} and filtered out any that failed our atom typing scheme, yielding 7,559,852 distinct small molecules. Up to 20 conformers were generated for each compound using RDKit, with an average of 14 conformations per compound. Out of the 108,878,042 resulting molecular structures, we randomly held out 10\% in a validation set for model optimization, while the remaining examples were used for training.

Models were evaluated on an independent test set of compounds collected from PubChem \cite{pubchem2019} with varying degrees of similarity to the training set. Any molecules that had more than one disconnected fragment, had molecular weight > 1,000 \si{\dalton}, or had elements not found in our atom typing scheme were excluded. The remaining compounds were then sorted into bins based on their maximum similarity to any molecule in the training set, using the Tanimoto coefficient between OpenBabel FP2 fingerprints. One thousand molecules were retained in each similarity bin of width 0.1 from 0.1 to 1.0, and 14 compounds were found with zero similarity to any training set molecule. A single conformer was generated for each test set molecule using RDKit.

\begin{figure}
    \centering
    \includegraphics[width=1.0\textwidth]{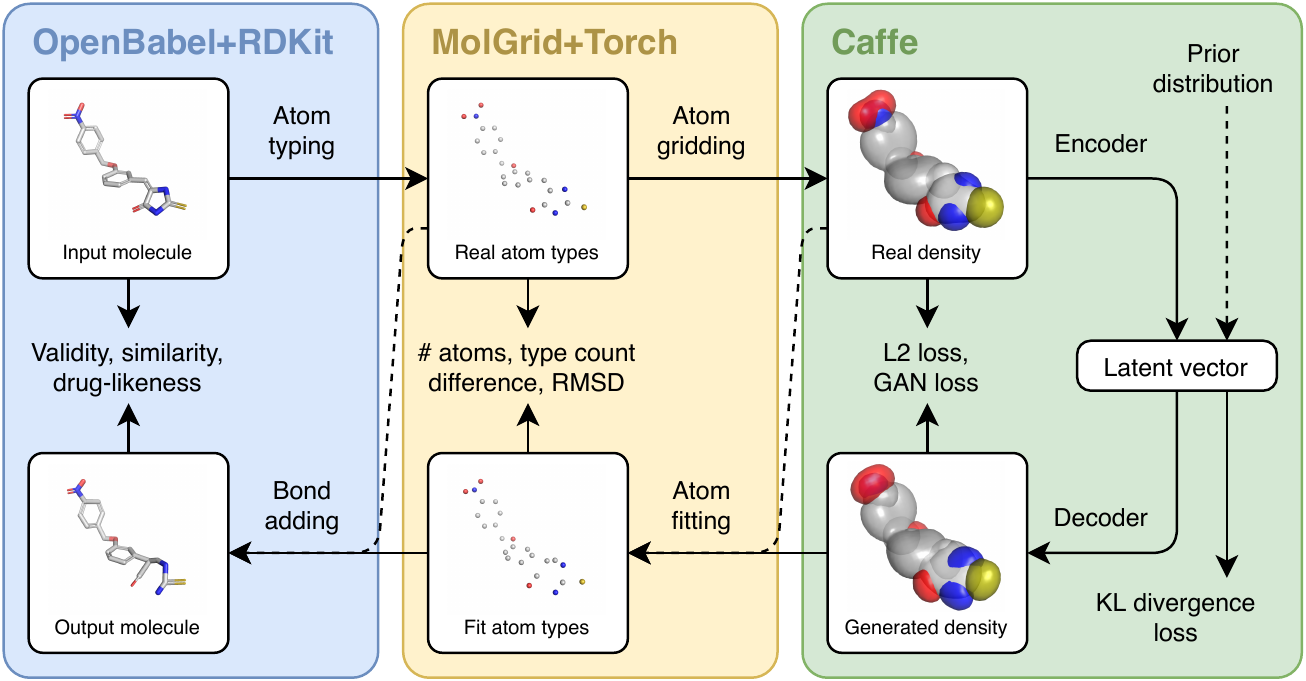}
    \caption{Overview of methods described in this work. Molecules are first converted to atom types and coordinates, then to atomic density grids. A generative model encodes real density to a latent space, then decodes to generated density. Atom fitting converts density to atom types and coordinates, then bond adding produces output molecules. Dashed lines indicate alternate pathways through the pipeline. From left to right, these are: adding bonds to real atom types and coordinates, fitting atoms to real densities, and decoding latent vectors from the prior distribution instead of from the encoder.}
    \label{fig:gen_methods}
\end{figure}

\subsection{Evaluation procedure}

Our general approach consisted of a pipeline through successive molecular representations, as depicted in Figure~\ref{fig:gen_methods}. The main pathway through the pipeline involved first typing and gridding real molecules to produce real densities. Next these were encoded into latent vectors using either the standard or variational autoencoder, then decoded as generated densities (AE posterior, VAE posterior). Densities were also generated by decoding latent vectors from a standard normal distribution with the variational autoencoder (VAE prior). Atom fitting was applied to generated densities to produce atom types and coordinates, followed by bond adding to yield valid output molecules.

Alternate pathways through our pipeline served as baselines that allowed us to assess the performance of each stage in isolation. We fitted atom types and coordinates to real density grids in order to evaluate atom fitting independently from the generative models (Real density). This also established an upper bound on our expected performance in atom-level metrics. We did the same for evaluating bond adding and molecule-level metrics by adding bonds to real atom types (Real atom types).

We employed a variety of metrics based on the different representations to evaluate our methods. We used L2 loss to directly compare generated densities to real densities and assess convergence while training our neural networks. After fitting atoms to real or generated densities, we considered atom type count differences and RMSD with respect to the real atom types and coordinates. Once molecules were constructed by adding bonds to atomic structures, we computed their validity, molecular weight, quantitative estimate of drug-likeness score (or QED score, in which higher is more drug-like) \cite{bickerton2012qed}, synthetic accessibility score (or SA score, in which lower is more synthesizable) \cite{ertl2009sas} , and Tanimoto similarity to the real input molecule using OpenBabel FP2 fingerprints.

During training, we evaluated each method on 1,000 random molecules from the validation set every 10,000 iterations. We generated output molecules for 10 different random rotations, and in the case of the VAE, different latent samples, of each input molecule. For the VAE prior, which was not conditioned on an input molecule, this simply meant generating 1,000 batches of 10 molecules from different latent samples. Once training converged, we evaluated each method on the different test set similarity bins. Again, we generated output molecules for 10 different random rotations and/or latent samples of each input molecule, or an equivalent number batches from the VAE prior.

We tested our ability to recover input atom types and coordinates from real densities and generated densities using atom fitting. After that we measured the overall validity of the molecules made by assigning bonds to the atomic structures from each method. We then evaluated the distributions of properties that are relevant to drug development on the valid generated molecules. Finally, we visualized some examples of sampling and interpolating between 3D molecules in the latent spaces learned by our generative models.

\begin{figure}
    \centering
    \includegraphics[width=\textwidth]{train_fit_metrics2.pdf}
    \caption{Molecular reconstruction, size, validity, and drug-likeness metrics on the validation set with respect to the training iteration. The filled-in areas show the average min and max across ten random samples of each input molecule or prior batch. }
    \label{fig:training_plots}
\end{figure}

\section{Results}

\subsection{Training convergence}

Figure~\ref{fig:training_plots} shows the training dynamics of our generative models on the validation set. The improving quality of generated densities as measured by L2 loss also resulted in better atom fitting. This is seen in the decreasing number of fit atom type differences relative to the input structure as training proceeded. By the end of training, the structures fit to AE densities had the exact input types with 68\% probability, and at least one sample out of 10 had the exact input types for 98\% of validation set molecules. Exact atom types were recovered from VAE posterior densities 15\% of the time overall, and at least one sample out of 10 had the exact atom types for 61\% of validation molecules.

In all generative modeling methods, the number of fit atoms increased with training. Molecules generated from the AE posterior converged to the average size of molecules in the true data distribution. The VAE posterior tended to produce slightly smaller molecules than the AE, while the prior molecules were significantly smaller than any of the other methods.

The validity shown in Figure~\ref{fig:training_plots} uses unmodified versions of OpenBabel \texttt{ConnectTheDots} and \texttt{PerceiveBondOrders} functions on the fit atomic coordinates instead of our full bond adding routine. These do not connect distant fragments with long bonds or use atom types to constrain bond adding, so this metric can be considered a simple proxy for geometric validity of the coordinates produced by atom fitting. Both the AE and VAE posterior molecules tended to produce increasingly valid coordinates over time, while the prior molecules had low coordinate validity for most of training.

The AE and VAE posterior generated molecules increasingly similar to the input as training progressed. After 100,000 iterations, the AE produced molecules with a Tanimoto similarity of 0.61 on average, and in a batch of 10 samples the most similar molecule had an expected similarity of 0.88. Around 54\% of the validation set molecules were perfectly reconstructed in at least one sample from the AE. The VAE posterior molecules had an average similarity of 0.32 to their input molecule and less than 10\% yielded the same exact molecule in any sample.

The QED score of valid molecules created by the AE approached the mean of the validation set by iteration 100,000. The drug-likeness of the VAE posterior increased during training as well. For both the AE and VAE posterior, the most drug-like output sample had a higher mean QED score than the input molecule. The QED scores of prior molecules converged lower than the posterior molecules, but the maximum QED score per batch was in the range of the posterior molecules.

% Rest of plots are on test sets

\subsection{Atom fitting reconstruction}

\begin{figure}
    \centering
    \includegraphics[width=\textwidth]{atom_fitting_recon.pdf}
    \caption{Ability to exactly fit atom types and coordinates of test set molecules to real densities or those generated from the autoencoder or variational autoencoder posterior, using 10 samples per molecule. In the line plots, the area from the expected sample min to the sample max is filled in.}
    \label{fig:atom_fitting_recon}
\end{figure}

Our capability to reconstruct atom types and coordinates of test set molecules from real and generated densities is shown in Figure~\ref{fig:atom_fitting_recon}. The mean L2 loss of AE and VAE posterior densities on the test set was around 14 and 27 respectively. There was little correlation with the test set similarity bin, suggesting that the neural networks were not overfit to the training set.

Fitting atoms to real densities produced an average type difference of 0.11 compared to the real atom types, and the exact input types were reconstructed with 98\% probability. The number of type differences from fitting to generated densities was 1.9 for the AE and 4.2 for the VAE posterior on the test set, which translated into recovering the exact types from 47\% of AE densities and 10\% of VAE posterior densities. The exact input types were recovered in at least one sample for 70\% of test set molecules using the AE and 34\% with the VAE posterior. Atom type differences were higher on generated densities when input molecules were less similar to the training set.

Whenever the input atom types were recovered exactly by atom fitting, we computed an RMSD to the real atoms by finding the minimum-distance assignment between atoms of the same type. The average RMSD was 0.011 \si{\angstrom} for atoms fit to real densities, 0.28 \si{\angstrom} when fitting to AE densities, and 0.47 \si{\angstrom} when fitting to VAE posterior densities. This indicates nearly identical coordinates in every case, with slightly more variation from the VAE.

\subsection{Molecular validity}

Once atom fitting was validated, we assessed the ability to produce valid molecules by adding bonds to atom types and coordinates from each method, using the test set as input. We were able to produce valid molecules from over 99\% of real atom types and atoms fit to real densities. Using our generative models, we created valid molecules from 92\% of AE densities, 91\% of VAE posterior densities, and 91\% of VAE prior densities. At least one valid sample was generated for 98\% of test set molecules using the AE, 99\% using the VAE posterior, and for 100\% of prior batches. The most common reason for invalid molecules was an atom having a greater valence than permitted, followed by failing to assign discrete bond orders to ring systems marked aromatic.

\subsection{Properties of valid molecules}

\begin{figure}
    \centering
    \includegraphics[width=\textwidth]{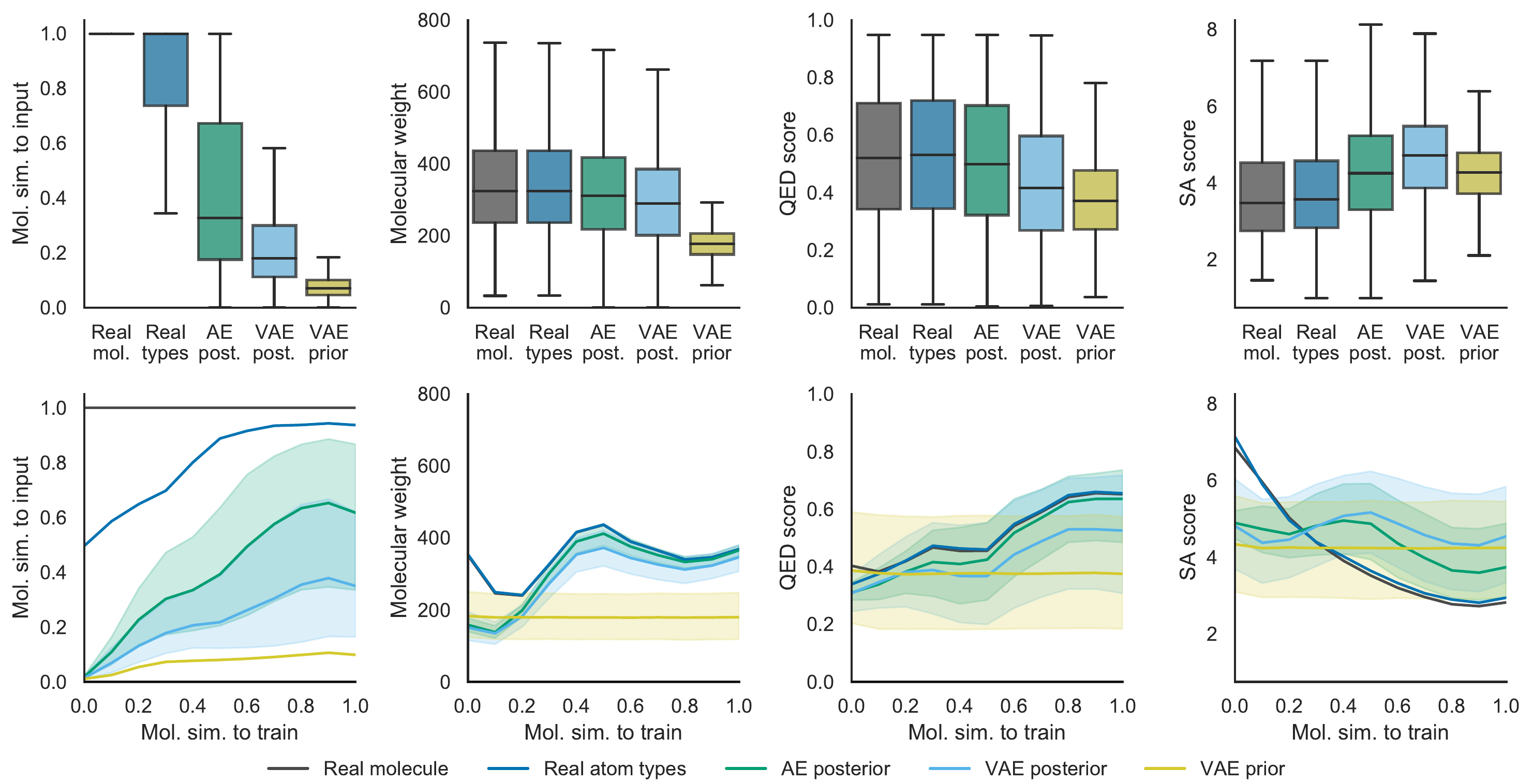}
    \caption{Properties of valid molecules from the test set (Real mol.), from bond adding to real atom types (Real types), and from the generative models (AE post., VAE post., VAE prior). Ten output samples were produced for each input molecule, and the range from the expected sample min to sample max is filled in on the line plots.}
    \label{fig:mol_properties}
\end{figure}

Figure~\ref{fig:mol_properties} shows drug property distributions of valid molecules from our generative models on the test set compared with baseline methods. In addition to using the real test set molecules as a baseline, we included molecules from adding bonds to real atom types and coordinates. The output molecules from real atom types had mean Tanimoto similarity greater than 0.82, and a median of 1.0, to the input molecule. The similarity to the input declined as the input molecules became less similar to the training set. Despite this, the distributions of molecular weight, QED score, and synthetic accessibility of molecules from adding bonds to real atoms were all highly matched with the real test set molecules.

The expected Tanimoto similarity of AE posterior molecules to their input was 0.43, and the most similar output sample was around 0.63. VAE posterior molecules had a lower expected similarity of 0.24, with the most similar sample expected at 0.43. The similarity of VAE prior molecules to random test set molecules was only 0.13. The output molecules from every generative method decreased in similarity to the input as it became less similar to the training set, following the same trend seen in bond adding to real atoms.

The molecules generated from the AE and VAE posterior were similar in size to real molecules by molecular weight, though slightly smaller on average. VAE prior molecules were significantly smaller than either VAE posterior or real molecules. When the input molecule had less than 0.3 similarity to the training set, the size of the output molecules from the AE and VAE posterior fell to around that of the prior distribution. On all other test set bins, the molecular weight of posterior molecules was highly correlated with the size of the input.

The mean QED score of AE posterior samples was 0.49, which is near the drug-likeness of real molecules in our test set. The VAE posterior and prior molecules had slightly lower expected QED scores of around 0.43 and 0.38. The most drug-like output molecule in ten samples had higher expected QED than the input molecule, using either the AE or VAE posterior. This trend held across every test set bin except molecules with 0.0 similarity to the training set. The SA score of molecules from the generative models tended slightly higher than real molecules, indicating lower synthesizability. However, they were more synthesizable than the most dissimilar test molecules.

\subsection{Latent sampling and interpolation}

\begin{figure}
    \centering
    \includegraphics[width=\textwidth]{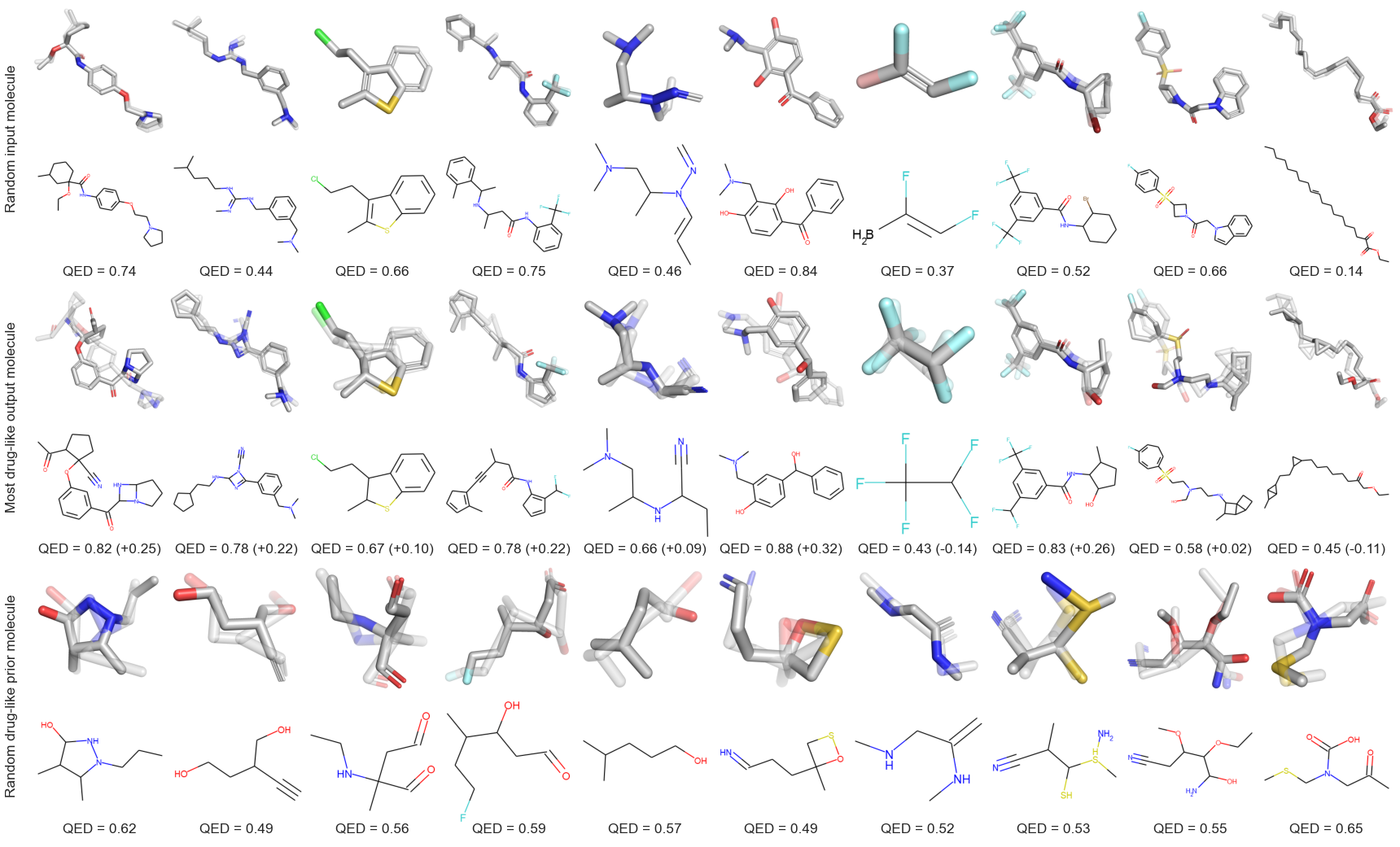}
    \caption{Random selection of test set molecules, the most drug-like output molecules in 10 VAE posterior samples using the above molecule as input, and the most drug-like output molecules from unrelated batches of 10 VAE prior samples. Each molecule is annotated with its QED score.}
    \label{fig:vae_post_mols}
\end{figure}

Finally, we exemplify two potential use cases of the continuous latent spaces in which our generative models learned to represent 3D molecules. Figure~\ref{fig:vae_post_mols} shows examples of molecular structures produced by drawing 10 random samples from either the VAE posterior or prior and selecting the sample with the highest QED score. Each molecule is displayed in 2D and 3D, with the 3D structure minimized using UFF \cite{rappe1992uff}. The unminimized 3D conformation produced by the generative model is overlaid with transparency. 80\% of the VAE posterior examples shown have higher QED score than their input. The amount that the VAE posterior outputs were modified compared to the input varied widely--some examples differed in only a single bond or functional group, while others reconfigured the entire molecule. There was a clear tendency to alter rings and scaffolds in unusual ways, creating structures that bear a high degree of shape similarity but little topological resemblance. On the other hand, the VAE prior molecules were mostly small, fragment-like structures with numerous polar functional groups but not much scaffold diversity.

\begin{figure}
    \centering
    \includegraphics[width=\textwidth]{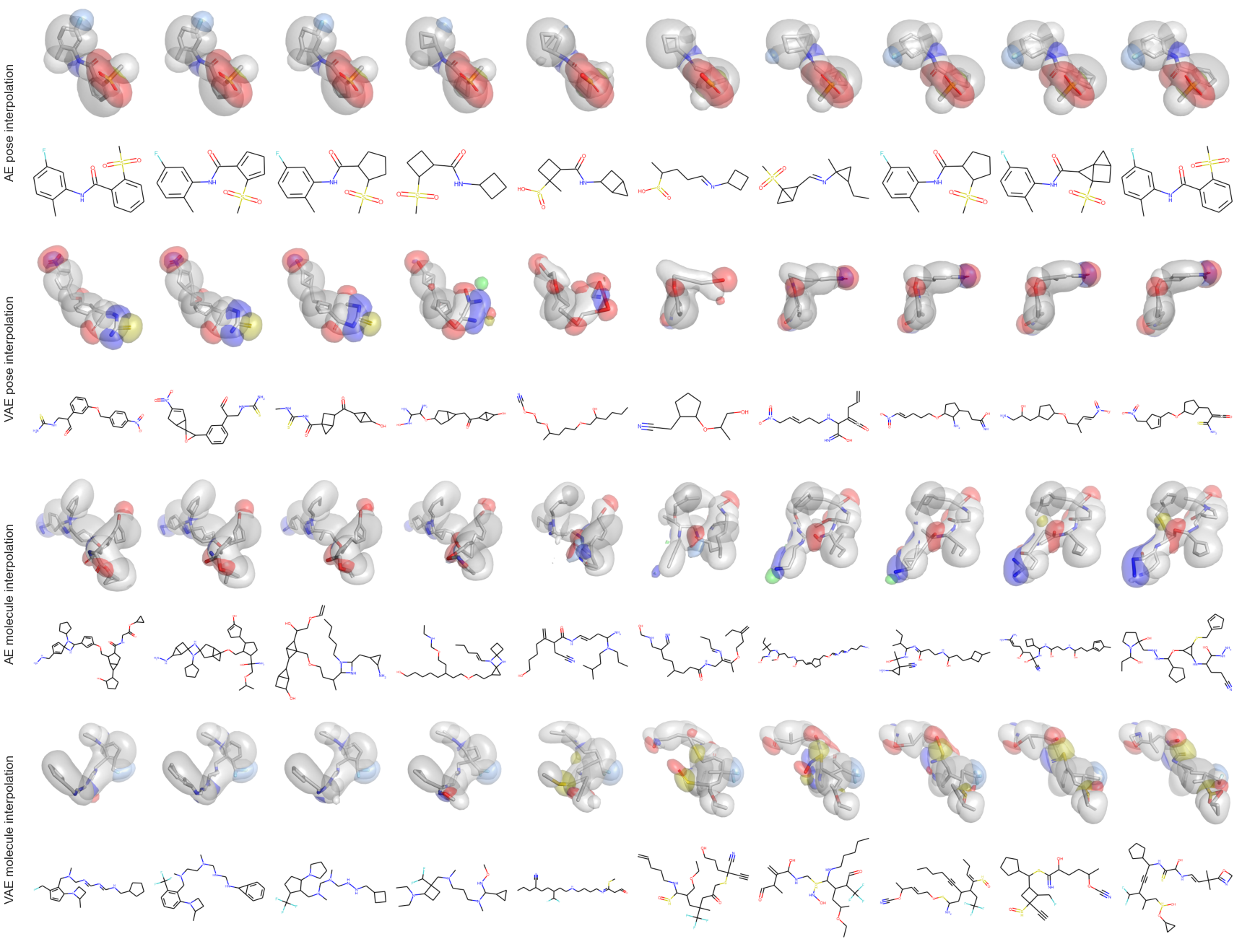}
    \caption{Four different interpolations between molecules in latent space, each shown in 3D with overlaid structures and densities and also as 2D structures. No minimization was performed on the structures. The first interpolates between two poses of the same molecule using the AE. The second interpolates between two samples of the same molecule using the VAE. The third and fourth use the AE and VAE to interpolate between two entirely different molecules.}
    \label{fig:latent_interp}
\end{figure}

In Figure~\ref{fig:latent_interp}, four different interpolations were performed between different molecules or different poses of the same molecules through the latent spaces learned by our generative models. The AE interpolations were performed linearly, while the VAE interpolations used spherical linear interpolation. There is some degree of chemical structure maintained throughout the interpolations. For instance, the sulfonyl and amide groups are present in almost every step of the AE pose transition. Most of the transformations maintain overall shape continuity rather than topological or chemical similarity. This is evidenced by aromatic densities that smoothly change size and position translating into aromatic rings that become smaller, non-aromatic rings or alkane branches.

\section{Discussion}

By training deep generative models on atomic density grids and applying our atom fitting and bond adding algorithms, we have developed the first deep learning model for generating three-dimensional molecular structures with diverse chemical compositions. Our models generate valid molecules at a rate greater than 90\%, but we realized that validity may be an insufficient metric for assessing generative models of 3D molecules since it ignores geometry and energetic favorability. Molecules that were made valid by adding strained bonds to ensure a single connected fragment often had high energy. This highlights the need for different evaluation criteria for generating 3D molecules than those used for 2D molecules.

Our standard AE was successful at exactly reconstructing its input atom types and coordinates in about half of test set densities, but fell off when evaluating molecules that were increasingly dissimilar to the training set, as seen in Figure~\ref{fig:atom_fitting_recon}. This was not attributable to atom fitting, since we were able to reconstruct the exact atom types in 98\% of real densities regardless of the test set bin and attain extremely low RMSD. On the contrary, the disparity in fitting to generated densities was reflected in the L2 loss of those densities, implying that future work should focus on improving the quality of the underlying generative models. It could also be a sign of overfitting and the need for a stricter validation set, since this was not detected in Figure~\ref{fig:training_plots}.

The VAE posterior was less likely to reconstruct the input, producing outputs that differed by about 4 atom types on average. This was expected as the VAE should, by design, produce variations of the input, and could be used to propose new drug-like molecular structures related to a given molecule. This was supported by the fact that the maximum QED score of the VAE posterior samples were higher than the QED of the input, as seen in Figure~\ref{fig:mol_properties}. Furthermore, there was no change in the synthetic accessibility of the VAE posterior molecules as the input became less similar to the training set. These observations suggest that the variational posterior had the capability of reliably proposing novel drug-like compounds based on input molecules.

The VAE prior molecules had slightly lower drug-likeness and higher synthetic accessibility than the posterior, in part due to their lower molecular weight. A primary objective in variational models is gaining the ability to sample useful outputs from the prior, and in their current state our VAE prior molecules are too small and structurally simple. Moving forward, an increased emphasis will be placed on transferring the relative success of posterior sampling to the prior, for instance by increasing the KL divergence loss weight over time or explicitly training on prior samples.

Given that we generated molecules with a Tanimoto similarity of only 0.82 to the real molecule from real atom types and coordinates, there is room for improvement in bond adding. This is likely due to limitations in our atom typing scheme, which lacks the information (e.g. formal charges, hybridization states) necessary to perfectly reconstruct certain molecules. An alternative atom typing scheme that provides better information for bond adding could improve our molecular reconstruction accuracy and the diversity of our novel generated molecules. 

This work serves as a proof of concept for generating 3D molecules using deep learning that can be expanded in numerous ways. The ability to map molecular structures to and from a continuous representation introduces the possibility of applying continuous optimization to the generation of desirable molecular structures. While 2D generative models have utilized cheminformatic property optimization, we are currently integrating protein structures into our pipeline to create 3D structure-conditional generative models. These are tasked with learning distributions over bound ligand structures conditioned on a binding pocket. The potential for new directions in 3D generative models for drug discovery are vast, and we hope that their exploration can be accelerated by making this project publicly available at \texttt{https://github.com/mattragoza/liGAN} and \texttt{https://github.com/gnina}.

%\begin{comment}
\section*{Acknowledgements}
 This work is supported by R01GM108340 from the National Institute of General Medical Sciences, is supported in part by the University of Pittsburgh Center for Research Computing through the resources provided, and used the Extreme Science and Engineering Discovery Environment (XSEDE), which is supported by National Science Foundation grant number ACI-1548562 through the Bridges GPU-AI resource allocation TG-MCB190049.
 %\end{comment}
 
\small
\bibliography{biblio}

\begin{thebibliography}{10}

\bibitem{Seifert2003}
Markus H.~J. Seifert, Kristina Wolf, and Daniel Vitt.
\newblock {Virtual high-throughput in silico screening}.
\newblock {\em BIOSILICO}, 1(4):143--149, Sep 2003.

\bibitem{sunseri2016pharmit}
Jocelyn Sunseri and David~Ryan Koes.
\newblock Pharmit: interactive exploration of chemical space.
\newblock {\em Nucleic acids research}, 44(W1):W442--W448, 2016.

\bibitem{krizhevsky2012imagenet}
Alex Krizhevsky, Ilya Sutskever, and Geoffrey~E Hinton.
\newblock {ImageNet Classification with Deep Convolutional Neural Networks}.
\newblock In {\em Advances in Neural Information Processing Systems}, pages
  1097--1105, 2012.

\bibitem{lecun2015deep}
Yann LeCun, Yoshua Bengio, and Geoffrey Hinton.
\newblock {Deep Learning}.
\newblock {\em Nature}, 521(7553):436--444, 2015.

\bibitem{angermueller2016dl}
Christof Angermueller, Tanel Pärnamaa, Leopold Parts, and Oliver Stegle.
\newblock {Deep learning for Computational Biology}.
\newblock {\em Mol. Syst. Bio.}, 12(878), 2016.

\bibitem{wallach2015atomnet}
Izhar Wallach, Michael Dzamba, and Abraham Heifets.
\newblock {AtomNet: A Deep Convolutional Neural Network for Bioactivity
  Prediction in Structure-Based Drug Discovery}.
\newblock arXiv preprint:1510.02855, 2015.

\bibitem{ragoza2017cnn}
Matthew Ragoza, Joshua Hochuli, Elisa Idrobo, Jocelyn Sunseri, and David~Ryan
  Koes.
\newblock {Protein-Ligand Scoring with Convolutional Neural Networks}.
\newblock {\em J Chem Inf Model}, 57(4):942--957, 2017.
\newblock
  [PubMed:\href{https://www.ncbi.nlm.nih.gov/pubmed/28368587}{28368587}]
  [doi:\href{http://dx.doi.org/10.1021/acs.jcim.6b00740}{10.1021/acs.jcim.6b00740}].

\bibitem{gomes2017}
Joseph Gomes, Bharath Ramsundar, Evan~N. Feinberg, and Vijay~S. Pande.
\newblock {Atomic Convolutional Networks for Predicting Protein-Ligand Binding
  Affinity}.
\newblock {\em arXiv preprint arXiv:1703.10603}, 2017.

\bibitem{jimenez2018kdeep}
José Jiménez, Miha Škalič, Gerard Martínez-Rosell, and Gianni
  De~Fabritiis.
\newblock {KDEEP: Protein–Ligand Absolute Binding Affinity Prediction via
  3D-Convolutional Neural Networks}.
\newblock {\em Journal of Chemical Information and Modeling}, 58(2):287--296,
  2018.
\newblock PMID: 29309725.

\bibitem{schutt2017}
Kristof~T. Schütt, Pieter-Jan Kindermans, Huziel~E. Sauceda, Stefan Chmiela,
  Alexandre Tkatchenko, and Klaus-Robert Müller.
\newblock {SchNet: A continuous-filter convolutional neural network for
  modeling quantum interactions}.
\newblock {\em arXiv preprint arXiv:1706.08566}, 2017.

\bibitem{kingma2014vae}
Diederik~P Kingma and Max Welling.
\newblock {Auto-Encoding Variational Bayes}.
\newblock {\em arXiv preprint arXiv:1312.6114}, 2014.

\bibitem{goodfellow2014gan}
Ian Goodfellow, Jean Pouget-Abadie, Mehdi Mirza, Bing Xu, David Warde-Farley,
  Sherjil Ozair, Aaron Courville, and Yoshua Bengio.
\newblock {Generative Adversarial Networks}.
\newblock {\em arXiv preprint arXiv:1406.2661}, 2014.

\bibitem{larsen2016vaegan}
Anders Boesen~Lindbo Larsen, Søren~Kaae Sønderby, Hugo Larochelle, and Ole
  Winther.
\newblock {Autoencoding beyond pixels using a learned similarity metric}.
\newblock {\em arXiv preprint arXiv:1512.09300}, 2016.

\bibitem{radford2016dcgan}
Soumith~Chintala Alec~Radford, Luke~Metz.
\newblock {Unsupervised Representation Learning with Deep Convolutional
  Generative Adversarial Networks}.
\newblock {\em arXiv preprint arXiv:1511.06434}, 2016.

\bibitem{gomez-bombarelli2016}
Rafael Gómez-Bombarelli, David Duvenaud, José~Miguel Hernández-Lobato, Jorge
  Aguilera-Iparraguirre, Timothy~D. Hirzel, Ryan~P. Adams, and Alán
  Aspuru-Guzik.
\newblock {Automatic chemical design using a data-driven continuous
  representation of molecules}.
\newblock {\em arXiv preprint arXiv:1610.02415v1}, 2016.

\bibitem{segler2017lm}
Marwin~H.S. Segler, Thierry Kogej, Christian Tyrchan, and Mark~P. Waller.
\newblock {Generating Focussed Molecule Libraries for Drug Discovery with
  Recurrent Neural Networks}.
\newblock {\em arXiv preprint arXiv:1701.01329}, 2017.

\bibitem{weininger1988smiles}
David Weininger.
\newblock {SMILES, a chemical language and information system. 1. Introduction
  to methodology and encoding rules}.
\newblock {\em Journal of Chemical Information and Computer Sciences},
  28(1):31--36, 1988.

\bibitem{kusner2017gramvae}
Matt~J. Kusner, Brooks Paige, and Jos{\'e}~Miguel Hern{\'a}ndez-Lobato.
\newblock {Grammar Variational Autoencoder}.
\newblock In Doina Precup and Yee~Whye Teh, editors, {\em Proceedings of the
  34th International Conference on Machine Learning}, volume~70 of {\em
  Proceedings of Machine Learning Research}, pages 1945--1954, International
  Convention Centre, Sydney, Australia, 06--11 Aug 2017. PMLR.

\bibitem{dai2018sdvae}
Hanjun Dai, Yingtao Tian, Bo~Dai, Steven Skiena, and Le~Song.
\newblock {Syntax-Directed Variational Autoencoder for Structured Data}.
\newblock In {\em International Conference on Learning Representations}, 2018.

\bibitem{guimaraes2018organ}
Gabriel~Lima Guimaraes, Benjamin Sanchez-Lengeling, Carlos Outeiral, Pedro
  Luis~Cunha Farias, and Alán Aspuru-Guzik.
\newblock {Objective-Reinforced Generative Adversarial Networks (ORGAN) for
  Sequence Generation Models}, 2018.

\bibitem{kearnes2016graphconv}
Steven Kearnes, Kevin McCloskey, Marc Berndl, Vijay Pande, and Patrick Riley.
\newblock {Molecular Graph Convolutions: Moving Beyond Fingerprints}.
\newblock {\em arXiv preprint arXiv:1603.00856}, 2016.

\bibitem{simonovsky2018graphvae}
Martin Simonovsky and Nikos Komodakis.
\newblock {GraphVAE: Towards Generation of Small Graphs Using Variational
  Autoencoders}.
\newblock {\em arXiv preprint arXiv:1802.03480v1}, 2018.

\bibitem{decao2018molgan}
Nicola De~Cao and Thomas Kipf.
\newblock {MolGAN: An implicit generative model for small molecular graphs}.
\newblock {\em arXiv preprint arXiv:1805.11973v1}, 2018.

\bibitem{jin2019jtvae}
Wengong Jin, Regina Barzilay, and Tommi Jaakkola.
\newblock {Junction Tree Variational Autoencoder for Molecular Graph
  Generation}.
\newblock {\em arXiv preprint arXiv:1802.04364v4}, 2019.

\bibitem{kwon2019graph}
Youngchun Kwon, Jiho Yoo, Youn-Suk Choi, Won-Joon Son, Dongseon Lee, and Seokho
  Kang.
\newblock {Efficient learning of non-autoregressive graph variational
  autoencoders for molecular graph generation}.
\newblock {\em J. Cheminform.}, 11(70), 2019.

\bibitem{gebauer2018distmat}
Niklas W.~A. Gebauer, Michael Gastegger, and Kristof~T. Schütt.
\newblock Generating equilibrium molecules with deep neural networks, 2018.

\bibitem{hoffmann2019distmat}
Moritz Hoffmann and Frank Noé.
\newblock Generating valid euclidean distance matrices, 2019.

\bibitem{skalic2019shapevae}
Miha Skalic, José Jiménez, Davide Sabbadin, and Gianni De~Fabritiis.
\newblock {Shape-Based Generative Modeling for de Novo Drug Design}.
\newblock {\em Journal of Chemical Information and Modeling}, 59(3):1205--1214,
  2019.
\newblock PMID: 30762364.

\bibitem{sunseri2020libmolgrid}
Jocelyn Sunseri and David~R Koes.
\newblock {libmolgrid: Graphics Processing Unit Accelerated Molecular Gridding
  for Deep Learning Applications}.
\newblock {\em J. Chem. Inf. Model.}, 60(3):1079--1084, 2020.

\bibitem{oboyle2011openbabel}
Noel~M. O'Boyle, Michael Banck, Craig~A. James, Chris Morley, Tim
  Vandermeersch, and Geoffrey~R. Hutchison.
\newblock {Open Babel: An open chemical toolbox}.
\newblock {\em J. Cheminf.}, 3(33), 2011.

\bibitem{openbabel}
{The Open Babel Package, version 3.1.1}.
\newblock \url{https://openbabel.org}.
\newblock (accessed May 29, 2020).

\bibitem{rdkit}
{RDKit: Open-Source Cheminformatics Software}.
\newblock \url{https://www.rdkit.org}.
\newblock (accessed May 28, 2020).

\bibitem{kingma2019adam}
Diederik~P. Kingma and Jimmy~Lei Ba.
\newblock {Adam: A Method for Stochastic Optimization}.
\newblock In {\em Proceedings of the 3rd International Conference Learning
  Representations}, 2015.

\bibitem{molport}
{MolPort: Easy compound ordering service}.
\newblock \url{https://www.molport.com}.
\newblock (accessed April 30, 2020).

\bibitem{pubchem2019}
Sunghwan Kim, Jie Chen, Tiejun Cheng, Asta Gindulyte, Jia He, Siqian He,
  Qingliang Li, Benjamin~A. Shoemaker, Paul~A. Thiessen, Bo~Yu, Leonid
  Zaslavsky, Jian Zhang, and Evan~E. Bolton.
\newblock {PubChem 2019 update: improved access to chemical data}.
\newblock {\em Nucleic Acids Res.}, 47:D1101--D1109, 2019.

\bibitem{bickerton2012qed}
Richard~G. Bickerton, Gaia~V. Paolini, Jérémy Besnard, Sorel Muresan, and
  Andrew~L. Hopkins.
\newblock {Quantifying the chemical beauty of drugs}.
\newblock {\em Nat Chem.}, 4(2):90--98, 2012.

\bibitem{ertl2009sas}
Peter Ertl and Ansgar Schuffenhauer.
\newblock {Estimation of synthetic accessibility score of drug-like molecules
  based on molecular complexity and fragment contributions}.
\newblock {\em J. Cheminform.}, 1(8):90--98, 2009.

\bibitem{rappe1992uff}
{Rappé, A. K. and Casewit, C. J. and Colwell, K. S. and Goddard III, W. A. and
  Skif, W. M.}
\newblock {UFF, a Full Periodic Table Force Field for Molecular Mechanics and
  Molecular Dynamics Simulations}.
\newblock {\em J. Am. Chem. Soc.}, 114:10024--10035, 1992.

\end{thebibliography}

\newpage

\section{Supplementary Materials}

\subsection{Model architecture details}

\begin{figure}[h]
    \centering
    \includegraphics[width=1.0\textwidth]{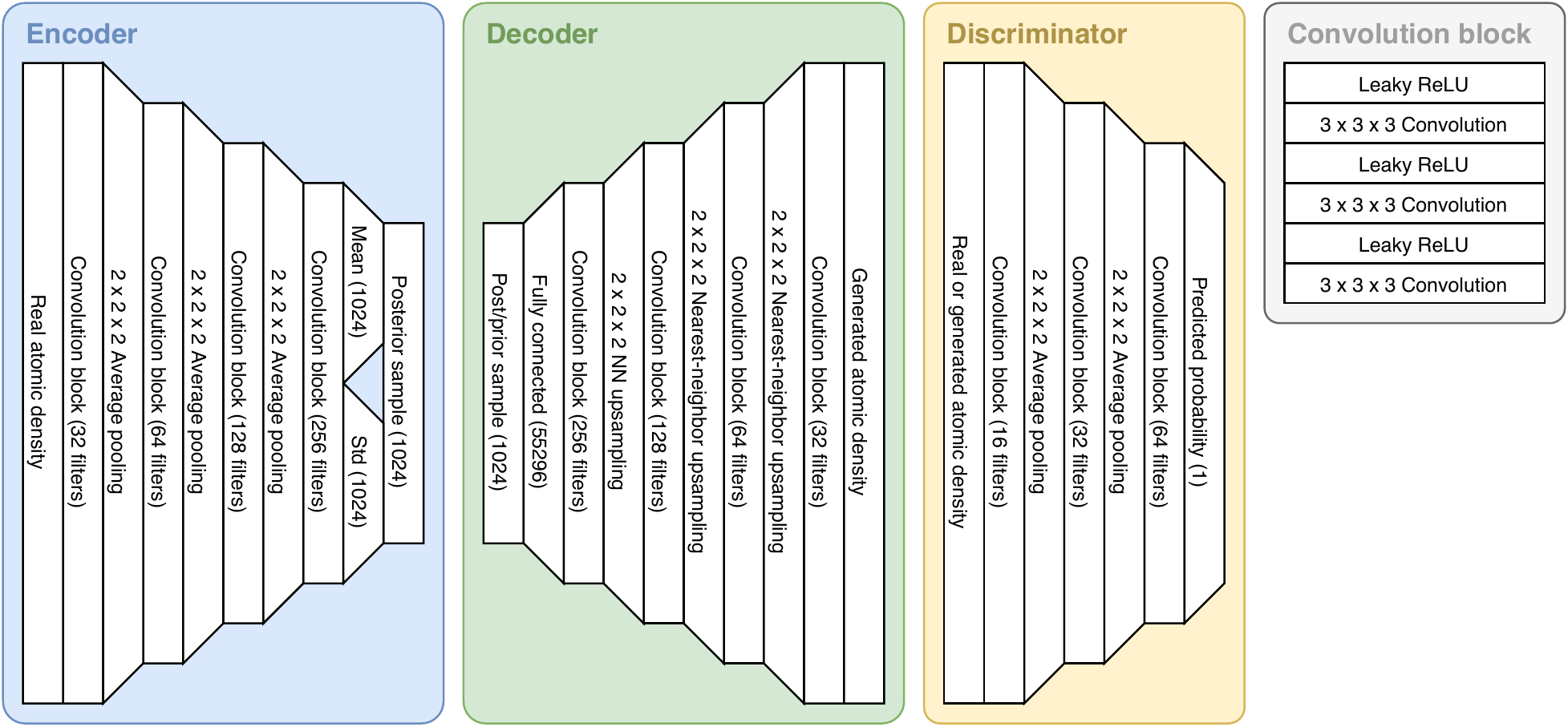}
    \caption{Generative and discriminative neural network architectures. The variational latent space is depicted, but the AE and VAE had identical architectures in every other respect. They consisted of symmetrical encoder and decoder sub-networks that were each a series of convolutional blocks interleaved with average pooling (in the encoder) or nearest-neighbor upsampling (in the decoder). Both the AE and VAE had a latent space of size 1024 and used leaky ReLU activation functions.}
    \label{fig:gen_model}
\end{figure}

\subsection{Atomic density and gridding functions}

In the molecular grid format, atoms are represented as continuous, Gaussian-like densities on a three-dimensional grid with separate channels for each atom type, analogous to an RGB image. The density of an atom at a grid point is a function of the displacement and the atomic radius:

\begin{align}
    \label{eqn:atom_density}
    f_t(c) =
    \begin{cases}
    e^{-2(\frac{\|c\|}{r_t})^2} & 0 \leq \|c\| < r_t \\
    \frac{4}{e^2}(\frac{\|c\|}{r_t})^2 - \frac{12}{e^2}\frac{\|c\|}{r_t} + \frac{9}{e^2} & r_t \leq \|c\| < \frac{3}{2}r_t \\
    0 & \frac{3}{2}r_t \leq \|c\|
    \end{cases}
\end{align}

\begin{figure}
    \centering
    \includegraphics[width=0.4\textwidth]{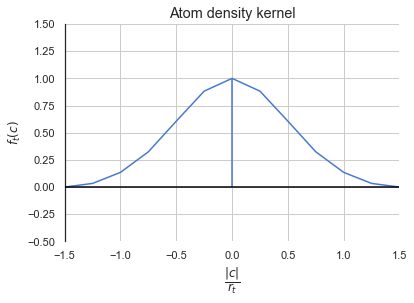}
    \caption{The Gaussian-like function, or kernel, from Equation~\ref{eqn:atom_density} that computes the density of a single atom at a grid point based on their displacement and the radius of the atom type.}
    \label{fig:atom_density_kernel}
\end{figure}

Given a 3D molecular structure with $N$ atoms and $N_T$ possible atom types as a vector of atom type indices $T \in [1, N_T]^N$ and a matrix of atomic coordinates $C \in \mathbb{R}^{N \times 3}$, the grid values in each atom type channel are computed by summing the density of each atom with the corresponding type at each grid point:

\begin{align}
    \label{eqn:atom_gridding}
    g(T, C)_{t,x,y,z} = \sum_{i=1}^{N} \mathbf{1}(T_i=t)  f_t(C_i - (x,y,z))
\end{align}

\subsection{Atom fitting algorithm}

\iffalse
\begin{align*}
T^\star, C^\star &= \argmin_{T, C} \mathcal{L}_{fit}(T, C) \\
\mathcal{L}_{fit}(T, C) &= \| \mathbf{G} - g(T, C) \|^2 \\
\mathcal{L}_{fit}(T, C) &= \sum_{t,x,y,z} (\mathbf{G} - g(T, C))^2 \\
\mathcal{L}_{fit}(T, C) &= \sum_{t,x,y,z} \mathbf{G}^2 - 2\mathbf{G}g(T, C) + g(T, C)^2 \\
\mathcal{L}_{fit}(T, C) &= \sum_{t,x,y,z} \mathbf{G}^2 - 2 \sum_{t,x,y,z} \mathbf{G}g(T, C) + \sum_{t,x,y,z} g(T, C)^2 \\
\mathcal{L}_{fit}(T, C) &= \|\mathbf{G}\|^2 + \|g(T, C)\|^2 - 2 \sum_{t,x,y,z} \mathbf{G}g(T, C) \\
\mathcal{L}_{fit}(T, C) &= \|\mathbf{G}\|^2 + \|g(T, C)\|^2 - 2 \sum_{t,x,y,z} \sum_{i=1}^{N} \mathbf{G} \mathbf{1}(T_i=t)  f_t(C_i - (x,y,z)) \\
\mathcal{L}_{fit}(T, C) &= \|\mathbf{G}\|^2 + \|g(T, C)\|^2 - 2 \sum_{t,x,y,z} \mathbf{G} \mathbf{1}(T=t)  f_t(C - (x,y,z)) \\
\mathcal{L}_{fit}(T, C) &= \|\mathbf{G}\|^2 + \|g(T, C)\|^2 - 2 \sum_{x,y,z} \mathbf{G}_T f_T(C - (x,y,z)) \\
\mathcal{L}_{fit}(T, C) &= \|\mathbf{G}\|^2 + \|g(T, C)\|^2 - 2 (\mathbf{G}_{T} *  f_T)(C) \\
T^\star, C^\star &= \underset{T, C}{\operatorname{argmin}} \|\mathbf{G}\|^2 + \|g(T, C)\|^2 - 2 (\mathbf{G}_{T} *  f_T)(C) \\
T^\star, C^\star &= \underset{T, C}{\operatorname{argmax}} \, (\mathbf{G}_{T} *  f_T)(C)
\end{align*}
\fi

\begin{algorithm}
    \small
    \DontPrintSemicolon
    \KwIn{An atomic density grid $\mathbf{G}_{ref} \in \mathbb{R}^{N_t \times N_x \times N_y \times N_z}$}
    \KwOut{The best-fit atom type indices $ T \in [1, N_t]^{N} $ and coordinates $ C \in \mathbb{R}^{N \times 3}$}
    \Begin
    {
        $T, C \gets (), () $\;
        $loss \gets \| \mathbf{G}_{ref} \|^2$ \;
        $struct_{init} \gets (loss, T, C)$\;
        $best\_structs \gets \{struct_{init}\} $\;
        $visited\_structs \gets \emptyset$ \;
        $found\_new\_best\_struct \gets true $\;
        \While{$found\_new\_best\_struct$}
        {
            $found\_new\_best\_struct \gets false $\;
            \For{$struct \in best\_structs$}
            {
                \If{$struct \in visited\_structs$}
                {   
                    continue \;
                }
                $loss, T, C \gets struct$\;
                $\mathbf{G}_{fit} \gets struct\_to\_grid(T, C) $\; 
                $\mathbf{G}_{diff} \gets \mathbf{G}_{ref} - \mathbf{G}_{fit} $\;
                \For{$(t_{new}, c_{new}) \in top\_k\_next\_atoms(\mathbf{G}_{diff})$}
                {
                    $T_{new} \gets$ append $t_{new}$ to $T$\;
                    $C_{new} \gets$ append $c_{new}$ to $C$\;
                    $loss_{new}, C_{new}  \gets gradient\_descent(\mathbf{G}_{ref}, T_{new}, C_{new}) $\;
                    $struct_{new} \gets (loss_{new}, T_{new}, C_{new})$\;
                    \If{$loss_{new} <$ any $struct \in best\_structs$}
                    {
                        Add $struct_{new}$ to $best\_structs$\;
                        $found\_new\_best\_struct \gets true $\;
                    }
                }
                Add $struct$ to $visited\_structs $\;
            }
            \If{$found\_new\_best\_struct$}
            {
                $best\_structs \gets top\_k\_structs\_by\_loss(best\_structs)$\;
            }
        }
        $loss, T, C \gets best\_struct\_by\_loss(best\_structs)$\;
    }
    \Return{$T, C$}\;
\caption{{\sc Atom Fitting} Find a set of atoms whose density minimizes the squared error w.r.t. the reference atomic density grid using beam search and gradient descent.}
\label{algo:atom_fitting}
\end{algorithm}

\begin{figure}[b]
    \centering
    \includegraphics[width=0.6\textwidth]{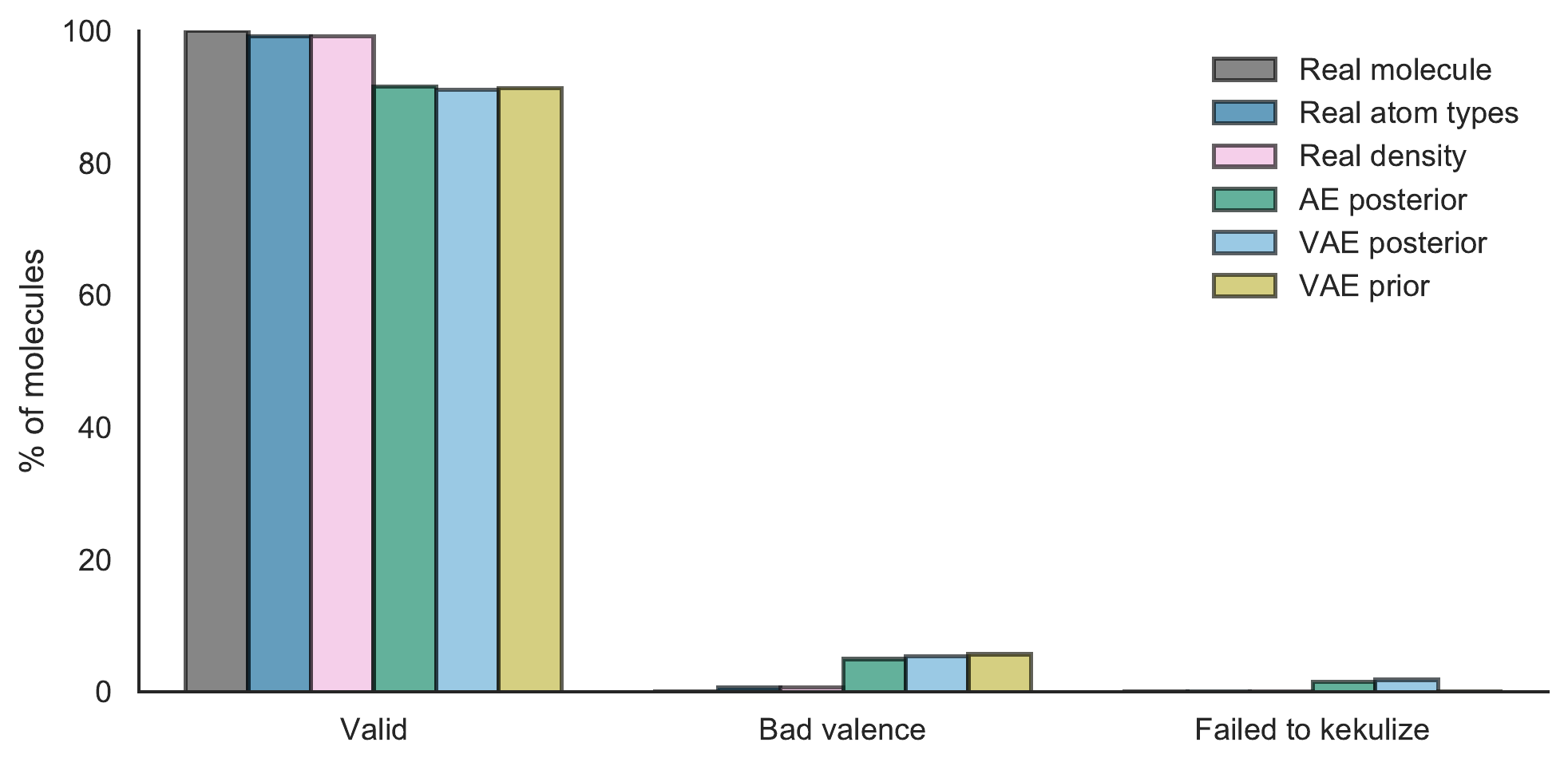}
    \caption{The percent of molecules from each method that are valid. The invalid molecules are broken down by the reason they failed sanitization in RDKit.}
    \label{fig:mol_validity}
\end{figure}

\end{document}